\documentclass[aps,pra,reprint]{revtex4-2}

\usepackage{lineno}
\usepackage{xcolor}

\usepackage{graphicx} 
\graphicspath{figures/}
\usepackage{xspace}

\usepackage{amsmath}  
\usepackage{amssymb}  
\usepackage{bm}       
\usepackage{mathrsfs} 
\usepackage{dcolumn}  
\usepackage{times}    
\usepackage{fancyhdr} 
\usepackage{multirow}

\usepackage[colorlinks,linkcolor=blue,anchorcolor=blue,citecolor=blue,urlcolor=blue]{hyperref}

\UseRawInputEncoding 
\begin{document}
\title{Loss-induced quantum nonreciprocity and entanglement in superconducting qubits}
\author{Yu-Meng Ren}
\author{Peng-Bo Li}
\email{lipengbo@mail.xjtu.edu.cn}
\affiliation{Ministry of Education Key Laboratory for Nonequilibrium Synthesis and Modulation of Condensed Matter, Shaanxi Province Key Laboratory of Quantum Information and Quantum Optoelectronic Devices, School of Physics, Xi'an Jiaotong University, Xi'an 710049, China}

\date{\today}

\begin{abstract}
Losses are ubiquitous in physics and are usually regarded as harmful in quantum information processing. Here, we propose a loss-induced scheme to achieve nonreciprocity and nonreciprocal entanglement in a superconducting platform, where two remote superconducting transmon qubits are connected via two lossy auxiliary cavities. The nonreciprocity in our scheme originates from interference between multiple lossy coupling paths. The coherent phases associated with the qubit-resonator couplings reverse sign under propagation reversal, while the loss-induced phases remain direction independent. Their combined effect leads to different interference conditions in the opposite directions, resulting in unequal effective couplings. We show that this loss-induced scheme can generate nonreciprocal quantum entanglement, indicating that loss can be utilized as a resource. Moreover, the tunability of nonreciprocity and nonreciprocal entanglement in our scheme can be manipulated by the relative phase induced by loss, allowing to tailor both reciprocal and nonreciprocal behaviors. Our results establish a direct link between engineered loss and nonreciprocal entanglement in quantum information processing and offer potential applications in scalable quantum networks.
\end{abstract}
\maketitle

\section{\label{sec:I}Introduction}
Nonreciprocity, which breaks the internal principle of time-reversal symmetry, plays a critical role in both classical and quantum regime~\cite{Yao2025,Avni2025,Brauns2024,Cotrufo2024,King2024,Zhang2023,Caloz2018}. It enables the unidirectional signal transmission, protecting signal sources from backscattered noise and suppressing unwanted modes~\cite{Wang2009,Malz2018}. Nonreciprocal devices, such as isolators~\cite{Shoji2018,Ren2022a} and circulators~\cite{Scheucher2016,Pintus2019}, typically rely on magneto-optical effects to ensure simultaneous transmission and reception on the same communication channel. However, the requirement of bulky components and strong magnetic fields in conventional magneto-optical schemes imposes restrictions on scalability and on-chip integration, especially in large-scale cryogenic quantum platforms. To overcome these challenges, considerable efforts have been devoted to developing magnetic-free nonreciprocal mechanisms, including the methods rely on spatiotemporal modulation~\cite{Sounas2017}, synthetic gauge fields~\cite{Shen2023}, optomechanical interactions~\cite{Ruesink2016,Peterson2017,Bernier2017,Fang2017,Xu2019},  nonlinearity~\cite{RosarioHamann2018,Yang2020,Tang2022,Xie2024,Zhou2024,Liu2025,Li2025,Wang2025} and cooperative effect of coherent and dissipative couplings~\cite{Metelmann2015,Wang2019,Ren2025}.
Beyond the classical transmission isolation, these approaches to achieve nonreciprocity also facilitate the exploration of nonreciprocal quantum effects, such as nonreciprocal photon and phonon blockade~\cite{Huang2018,Xu2020,Yao2022}, nonreciprocal quantum batteries~\cite{Ahmadi2024}, nonreciprocity in photon pair correlations~\cite{Graf2022}, nonreciprocity in critical phenomena~\cite{Begg2024} and nonreciprocal entanglement~\cite{Jiao2020,Ren2022,Guan2024}, offering new avenues for quantum information processing.

Loss has been regarded as a major obstacle in quantum information processing for a long time. Eliminating the adverse effects of loss and developing the strategies to utilize it as a resource is therefore significant. In non-Hermitian physics, loss is usually associated with the emergence of exceptional points (EPs), which are non-Hermitian spectral degeneracies characterized by the coalescence of both eigenvalues and eigenfunctions~\cite{Miri2019,Tschernig2022,Li2023}. By precisely tuning the loss parameters, the system can be driven into the vicinity of an EP, resulting in mode reorganizations and enabling unconventional effects such as loss-induced transparency~\cite{Guo2009,Zhang2018,Beder2024}, unidirectional invisibility~\cite{Regensburger2012}, the suppression and revival of lasing~\cite{Peng2014} and loss-enabled chirality inversion~\cite{He2025}. In the context of topological physics, loss can also play an active role in inducing nonreciprocal transport through the non-Hermitian skin effect~\cite{Li2020,Wu2025} and in facilitating the annihilation and revival of topological phase singularity pairs~\cite{Liu2021}. Alternatively, a distinct route to exploit the loss as a resource has emerged recently, facilitating the development of loss-induced nonreciprocity~\cite{Huang2021,Huang2023,Li2024}. Such systems with multichannel interference and direction-independent phase induced by loss, can break Lorentz reciprocity by behaving asymmetry between different propagation channels. Despite these exciting developments, existing studies either focus on the classical domain, exploring the nonreciprocal energy transmission~\cite{Huang2021}, or investigate the quantum correlations in boson systems~\cite{Li2024}. The loss-induced nonreciprocity and nonreciprocal quantum effects between remote qubits have been barely explored so far, especially concerning whether loss can be utilized to engineer nonreciprocal entanglement. 

Superconducting quantum platforms have recently attracted broad attention owing to the outstanding performance of superconducting qubits~\cite{Devoret2013,You2011,Kjaergaard2020}, which characterizes with enhanced coherence~\cite{Gyenis2021,Zhang2021,Somoroff2023,Milul2023}, flexible controllability~\cite{Xiang2013}, and tunable qubit couplings~\cite{Yan2018,Kounalakis2018,Li2020a,Xu2020a,Liu2025a}. These platforms are widely regarded as promising candidates for scalable quantum computation~\cite{Arute2019,Jurcevic2021,Wu2021,Krinner2022,Lin2022,Hangleiter2023,Ai2025,Deng2025,Song2025,Mollenhauer2025} and quantum simulation~\cite{Wendin2017,Gong2021,Shi2023,Zhang2023b,Rosen2025}. As a solid-state version analogy to cavity QED, circuit quantum electrodynamics (circuit QED) provides a paradigmatic architecture where superconducting qubits are coupled to resonators coherently~\cite{Blais2021,Clerk2020,Gu2017,Chiorescu2004}. This architecture has facilitated the exploration of a number of phenomena~\cite{Johansson2006,Houck2012,Stassi2013,Felicetti2014,Pechal2014,Paik2016,Kockum2017,Wang2017,Lin2019,Wang2020,Ao2023,Dakir2023,Liul2023,Kadijani2024,Maurya2024,Tomonaga2025}, extending from the dispersive regime to the strong coupling and even ultrastrong coupling domain~\cite{Wallraff2004,Schuster2007,Romero2012,Bosman2017,FornDiaz2019,PuertasMartinez2019}. Since both the strengths and detunings of qubit-resonator couplings can be precisely tuned~\cite{Zhou2021,Zeytinoifmmodegelsegfilu2015,Gambetta2011,Blais2007}, the qubit-resonator-qubit architecture naturally enable to mediate effective interactions and generate entanglement between superconducting qubits~\cite{Majer2007,DiCarlo2009,Poletto2012,Shankar2013,Quintana2013,Aron2016,Berrada2024,Kang2025}, offering a setting where engineered nonreciprocity is technically feasible. When the intermediate resonators are lossy, their adiabatic elimination yields complex-valued effective qubit-qubit couplings with controllable loss-induced phases, a mechanism that has been exploited to realize nonreciprocity between resonate bosonic modes~\cite{Huang2021}. In large-scale superconducting circuits, however, direction-selective couplings mainly rely on ferrite-based nonreciprocal elements for signal routing and back-action suppression. Although these devices offer high isolation, their magnetic materials and bulky sizes make them incompatible with dense cryogenic integration. The ability to develop on-chip, magnet-free and tunable nonreciprocal interactions within superconducting circuits becomes increasingly important.
Despite recent demonstrations of directional emission in superconducting platform through relative phases control and collective effects~\cite{Kannan2023,Mirhosseini2019,Loo2013}, the possibility to utilize loss of connecting modes as a resource to engineer nonreciprocity between superconducting qubits remains largely unexplored.

Here, the scheme in this work consists of two superconducting qubits coupled via two common lossy resonators. By adiabatically eliminating the connecting modes, a tailored effective qubit-qubit interaction exhibits asymmetry excitation transfer between leftward and rightward directions, thus the unidirectional coupling between qubits can be engineered. Furthermore, we not only show that our loss-induced scheme achieves nonreciprocal transmission but also induces nonreciprocal entanglement between two superconducting qubits. In contrast to the previous superconducting implementations~\cite{Ren2025}, where nonreciprocity is achieved through a precise balance between coherent and dissipative couplings with the coherent coupling established via a direct capacitor, our scheme in this work adopts a different configuration. In this work, two remote superconducting qubits are indirectly connected via two auxiliary resonator modes, enabling spatial separation between qubits. This mediated and remote configuration naturally supports scalability: it enables long-range qubit-qubit interactions while strongly suppressing crosstalk which typically arises between neighboring qubits, thereby avoiding the need for calibration or other complex protocols to reduce crosstalk noise. It is also suitable to support modular architectures for quantum computation, where unidirectional interactions can be essential for controlled information flow between processor units. Such long-range coupling is also in line with the growing experimental trend in circuit QED towards modular quantum processors and integration~\cite{Almanakly2025,Qiu2025}. Importantly, the nonreciprocity achieved in our scheme originates from the interference among multiple lossy coupling paths, without requiring strong magnetic fields, nonlinearity or spatially adjacent with direct coupling. This makes our architecture well-suited for the realization of on-chip nonreciprocal elements for scalable superconducting quantum networks. Therefore, this work provides a new method where engineered loss serves as a resource to achieve nonreciprocity in superconducting platform, revealing new opportunities for quantum information processing.

\section{\label{sec:II}The model}
\textcolor[rgb]{0,0,0}{In this section, we first introduce the bare Hamiltonian in the physical setup and present the corresponding engineered setup where two qubits are coupled to two lossy connecting modes through complex-valued couplings. Then we show how the parametric frequency modulation generates these engineered complex couplings. Finally, using the resulting engineered Hamiltonian and open-system dynamics, we derive the conditions for nonreciprocity originating from the interference of multiple lossy coupling channels.}
\begin{color}[rgb]{0,0,0}
\subsection{\label{sec:IIA}System setup and Hamiltonian}
\begin{figure*}[t]
	\centering
	\includegraphics[width=0.9\linewidth]{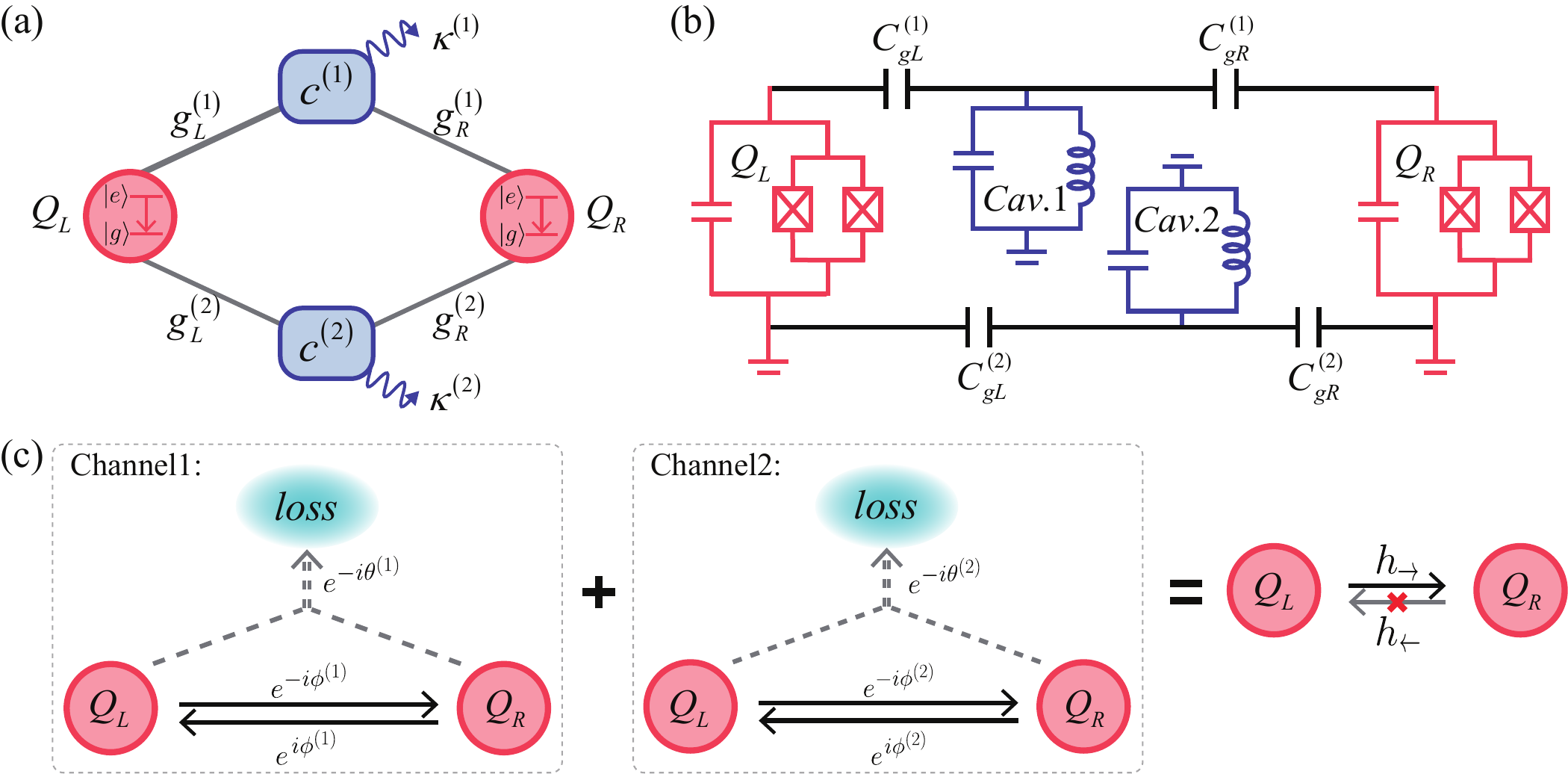}
	\caption{\textcolor[rgb]{0,0,0}{(a) Schematic of the engineered system model. A pair of two qubits $Q_{L}$ and $Q_{R}$ are connected by a series of connecting modes $\hat{c}^{(n)}$, with the engineered complex couplings $g^{(n)}_{L/R}$. These engineered couplings are generated by applying parametric modulations to the qubits in the presence of the bare couplings.}
	(b) Circuit implementation of two transmon qubits (pink) $Q_{L}$ and $Q_{R}$ connected by two auxiliary cavities (blue) via capacitive couplings. (c) Schematic diagram of the realization of nonreciprocity between qubits through the interfere between two channels. In each channel, the loss-induced phase factor $\theta^{(n)}$ are the same for leftward and rightward effective couplings. While the coherent factor $\phi^{(n)}$ contribute the opposite sign for different coupling directions.}
	\label{fig1:model}
\end{figure*}
As schematically depicted in Fig.~\ref{fig1:model}(a), we considered an engineered effective configuration where a pair of two qubits, $Q_{L}$ and $Q_{R}$, are connected via two lossy connecting modes $\hat{c}^{(1)}$ and $\hat{c}^{(2)}$, with the engineered complex couplings $g^{(n)}_{L/R}$. These couplings constitute the engineered model studied throughout the main text. To connect this effective description to a physical implementation, we start from the bare Hamiltonian in the laboratory frame($\hbar=1$):
\begin{equation}
	\begin{aligned}
		\hat{H} =&\sum_{m=L,R}\frac{1}{2}\omega_m(t)\hat{\sigma}_m^{z}+\sum_{n=1}^2 \omega_c^{(n)}\hat{c}^{(n)\dagger} \hat{c}^{(n)} \\
		+&\sum_{n=1}^2 \left[\left(\lambda_{L}^{(n)} \hat{\sigma}_{L}^+ +\lambda_{R}^{(n)} \hat{\sigma}_{R}^+\right)  \hat{c}^{(n)} + \text{H.c.} \right].
	\end{aligned}
	\label{H_tot}
\end{equation}
where $\hat{\sigma}_m^+$~$(\hat{\sigma}_m^-)$ is the raising (lowering) operator of qubit $Q_{m}$, and $\hat{c}^{(n)\dagger}$~$(\hat{c}^{(n)})$ is the creation (annihilation) operator of the $n$-th connecting mode. Here, the subscript $m=L,R$ represents the qubit, while the superscript $(n)$ denotes the connecting mode. The first term of $\hat{H}$ in Eq.~(\ref{H_tot}) describes the free Hamiltonian of qubits with the tunable qubit frequency $\omega_m$, while the second term represents the free Hamiltonian of connecting modes with the frequency $\omega_c^{(n)}$. The last term describes the interaction between qubits and connecting modes, where $\lambda_{L}^{(n)}$ and $\lambda_{R}^{(n)}$ correspond to the bare coupling coefficients between the $n$-th connecting mode and the left ($Q_{L}$) or the right ($Q_{R}$) qubit, respectively.
\end{color}

Such a system can be naturally implemented in the circuit QED architecture, where superconducting transmon qubits can serve as qubits, and the coplanar waveguide resonators or transmission line waveguides can support connecting modes, as illustrated in Fig.~\ref{fig1:model}(b). The frequency of a transmon qubit can be tuned by applying magnetic flux through a SQUID loop, which modulates the effective critical current of the Josephson junction. For simplicity, we focus on the case where two qubits are resonant at the same frequency, $\omega_L=\omega_R=\omega_0$, and the influence of qubit frequency detuning is discussed in Sec.~\ref{sec:VI}.

Here, in writing Eq.~(\ref{H_tot}), we have adopted the rotating-wave approximation (RWA) in the qubit-connecting mode interaction. This approximation is well justified in the parameter regime considered here. Specifically, the transmon qubit and resonator frequencies typically lie in the order of a few $\mathrm{GHz}$, while the bare coupling strengths $\lambda_{L/R}^{(n)}$ are at least one order of magnitude smaller, $\lambda_{L/R}^{(n)} \ll \omega_0,\omega_c^{(n)}$. As a result, the counter-rotating terms oscillate rapidly at frequencies $\sim \omega_0+\omega_c^{(n)}$ and average out on the timescales relevant to the dynamics of interest.

\begin{color}[rgb]{0,0,0}
\subsection{\label{sec:IIB}Parametric generation of the engineered couplings}
We now show a parametric modulation scheme to generates the independently engineered complex couplings $g_{L/R}^{(n)}$ required in Fig.~\ref{fig1:model}(a). Since their tunable amplitudes and phases are essential for the loss-induced nonreciprocity mechanism studied here. The detailed derivation of this engineering scheme is presented in Appendix~\ref{sec:appendixA}.

We consider a two-tone flux modulation to the qubit frequencies~\cite{Zhou2021}:
\begin{equation}
	\omega_m(t)=\omega_0+A_{m,1}\cos(\omega_{d1}t+\psi_{m,1})+A_{m,2}\cos(\omega_{d2}t+\psi_{m,2})
\end{equation}
where $m=L,R$ denotes the qubits and $\omega_0$ is the average qubit frequency. Here $A_{m,1}$ and $A_{m,2}$ denote the amplitudes of the modulation applied to $Q_m$, $\omega_{d1}$ and $\omega_{d2}$ are the corresponding modulation frequencies, and $\psi_{m,1}$ and $\psi_{m,2}$ are their phases. Starting from the laboratory-frame Hamiltonian with bare qubit-connecting mode couplings $\lambda_{m}^{(n)}$, one may do some rotating frame transformations, perform the sideband selections and keep only the required sidebands within the rotating-wave approximation. As detailed in Appendix~\ref{sec:appendixA}, this yields the engineered effective Hamiltonian given by:
\begin{equation}
	\hat{H}^\prime = - \sum_{n=1}^2 \Delta^{(n)} \hat{c}^{(n)\dagger} \hat{c}^{(n)} + \sum_{n=1}^2 \left[ \left( g_{L}^{(n)} \hat{\sigma}_{L}^+ + g_{R}^{(n)} \hat{\sigma}_{R}^+ \right) \hat{c}^{(n)} + \text{H.c.} \right]
	\label{H_prime}
\end{equation}
where $\Delta^{(n)}$ denotes the residual engineered detuning of the $n$-th connecting mode after sideband selection. Here, the engineered couplings take the form:
\begin{equation}
	\begin{aligned}
		g_{m}^{(1)}&=\lambda_{m}^{(1)} J_1\left(\frac{A_{m,1}}{\omega_{d1}}\right) J_0\left(\frac{A_{m,2}}{\omega_{d2}}\right) e^{i\tilde{\psi}_{m,1}}  \\
		g_{m}^{(2)}&=\lambda_{m}^{(2)} J_0\left(\frac{A_{m,1}}{\omega_{d1}}\right) J_1\left(\frac{A_{m,2}}{\omega_{d2}}\right) e^{i\tilde{\psi}_{m,2}}.
	\end{aligned}
\end{equation}
where $\tilde{\psi}_{m,1}$ and $\tilde{\psi}_{m,2}$ are determined solely by the modulation phases and selected sideband orders. This modulation provides a route to select sideband couplings to the two connecting modes, enabling independent control of the engineered coupling magnitudes and phases. 

\subsection{\label{sec:IIC}Open-system description and nonreciprocal condition}
With the engineered couplings established above, we next turn to the dynamics of the qubit-connecting mode system. Since each connecting mode experiences an energy decay characterized by rates $\kappa^{(1)}$ and $\kappa^{(2)}$, respectively, the system is governed by the Lindblad master equation~\cite{Scully1997,Breuer2002,Reitz2022}:
\end{color}
\begin{equation}
	\frac{d\hat{\rho}}{dt}=-i\left[\hat{H}^{\prime},\hat{\rho}\right]+\sum_{n=1}^{2}\kappa^{(n)}\mathcal{D}\left[\hat{c}^{(n)}\right]\hat{\rho}
	\label{master-eq}
\end{equation} 
with the superoperator $\mathcal{D}[\hat{O}] \hat{\rho}=\hat{O} \hat{\rho} \hat{O}^{\dagger}-\frac{1}{2} \hat{O}^{\dagger} \hat{O} \hat{\rho}-\frac{1}{2} \hat{\rho} \hat{O}^{\dagger} \hat{O}$. When performing the adiabatic elimination of the connecting modes, it is more convenient to adopt the Heisenberg-Langevin formalism. Within this framework, the evolution equations of the relevant operators can be written as:
\begin{subequations}
	\begin{align}
		\frac{d\hat{c}^{(1)}}{dt}& =\left(i\Delta^{(1)}-\frac{\kappa^{(1)}}{2}\right)\hat{c}^{(1)}-i\left(g_{L}^{*(1)}\hat{\sigma}_{L}^{-}+g_{R}^{*(1)}\hat{\sigma}_{R}^{-}\right)+\hat{F}^{(1)} \label{eq:a} \\
		\frac{d\hat{c}^{(2)}}{dt}& =\left(i\Delta^{(2)}-\frac{\kappa^{(2)}}{2}\right)\hat{c}^{(2)}-i\left(g_{L}^{*(2)}\hat{\sigma}_{L}^{-}+g_{R}^{*(2)}\hat{\sigma}_{R}^{-}\right)+\hat{F}^{(2)} \label{eq:b} \\
		\frac{d\hat{\sigma}_{L}^{-}}{dt}& =ig_{L}^{(1)}\hat{\sigma}_{L}^{z}\hat{c}^{(1)}+ig_{L}^{(2)}\hat{\sigma}_{L}^{z}\hat{c}^{(2)} \label{eq:c} \\
		\frac{d\hat{\sigma}_{R}^{-}}{dt}& 
		=ig_{R}^{(1)}\hat{\sigma}_{R}^{z}\hat{c}^{(1)}+ig_{R}^{(2)}\hat{\sigma}_{R}^{z}\hat{c}^{(2)} \label{eq:d}
	\end{align}
\end{subequations}
Here, the Langevin noise operators $\hat{F}^{(1)}$ and $\hat{F}^{(2)}$ denote the Langevin noise associated with the environment, introduced by the standard Heisenberg-Langevin treatment~\cite{Scully1997}. And the expectation values of these operators vanish, $\left\langle \hat{F}^{(n)} \right\rangle =0$.

From Eq.~(\ref{eq:a}) and (\ref{eq:b}), the dynamics of the $n$-th connecting mode relax on the timescale $\tau_c^{(n)}\sim1/\left|\Delta^{(n)}+i\frac{\kappa^{(n)}}{2}\right|$. In contrast, the qubit operators vary on the slower timescale $\tau_q^{(n)}\sim1/\left|g_{L/R}^{(n)}\right|$. When $\tau_c^{(n)} \ll \tau_q^{(n)}$, equivalently $\left|\Delta^{(n)}+i\frac{\kappa^{(n)}}{2}\right|\gg\left(\left|g_{L}^{(n)}\right|,\left|g_{R}^{(n)}\right|\right)$, the connecting modes adiabatically follow the qubit operators, allowing us to set $d\hat{c}^{(n)}/dt=0$. Therefore, we obtain:
	\begin{equation} 
		\hat{c}^{(n)}=\frac{g_{L}^{*(n)}\hat{\sigma}_{L}^-+g_{R}^{*(n)}\hat{\sigma}_{R}^-}{\Delta^{(n)}+i\kappa^{(n)}/{2}}
		\label{eq:c_mode}.
	\end{equation}
After substituting it into Eq.~(\ref{eq:c}) and (\ref{eq:d}), the dynamics of qubit lowering operators can be decoupled from the connecting modes, yielding the effective equation of motion with the form:
\begin{equation}
	\begin{split}
		\frac{d\hat{\sigma}_{L}^{-}}{dt}=& -i\Lambda_{L}\hat{\sigma}_{L}^{-}+ih_{\leftarrow}\hat{\sigma}_{L}^{z}\hat{\sigma}_{R}^{-} \\
		\frac{d\hat{\sigma}_{R}^{-}}{dt}=& -i\Lambda_{R}\hat{\sigma}_{R}^{-}+ih_{\rightarrow}\hat{\sigma}_{R}^{z}\hat{\sigma}_{L}^{-}
		\end{split}
	\label{Dynamic_lowering}
\end{equation}
where $\Lambda_{L}$ and $\Lambda_{R}$ denote the self-influence due to the connecting modes:
\begin{equation}
	\Lambda_{L}= \sum_{n=1}^2 \frac{g_{L}^{(n)}g_{L}^{*(n)}}{\Delta^{(n)}+i\kappa^{(n)}/2},
	\Lambda_{R}= \sum_{n=1}^2 \frac{g_{R}^{(n)}g_{R}^{*(n)}}{\Delta^{(n)}+i\kappa^{(n)}/2}.
\end{equation}
While the second the first terms in Eq.~(\ref{Dynamic_lowering}) represent the influence from the other qubit. Specifically, the effective coupling coefficients
\begin{equation}
	h_{\leftarrow}=\sum_{n=1}^2 \frac{g_{L}^{(n)}g_{R}^{*(n)}}{\Delta^{(n)}+i\kappa^{(n)}/2},
	h_{\rightarrow}=\sum_{n=1}^2 \frac{g_{L}^{*(n)}g_{R}^{(n)}}{\Delta^{(n)}+i\kappa^{(n)}/2}
	\label{h_coefficients}
\end{equation}
describe the influence from the right-sided $Q_R$ to left-sided $Q_L$ and from $Q_L$ to $Q_R$, respectively. 

From Eq.~(\ref{h_coefficients}), it is obviously to note that the leftward and rightward effective coupling coefficients share a similar structure but differ due to the complex-value coupling and loss-induced asymmetry. Each connecting channel contribute to the effective coupling with amplitude and phase, which can be expressed as:
\begin{equation}
	\begin{split}
		h_{\leftarrow}=& \sum_{n=1}^2 G^{(n)} e^{i\phi^{(n)}-i\theta^{(n)}},\\
		h_{\rightarrow}=& \sum_{n=1}^2 G^{(n)} e^{-i\phi^{(n)}-i\theta^{(n)}}.
	\end{split}
	\label{h_landr}
\end{equation}
where the amplitude of the effective coupling coefficient for the $n$-th channel is defined as:
\begin{equation}
	G^{(n)} \equiv \frac{\left|g_{L}^{(n)}g_{R}^{*(n)}\right|}{\sqrt{(\Delta^{(n)})^2+(\kappa^{(n)}/2)^2}}.
\end{equation}
Here, the coherent phase factor $\phi^{(n)}\equiv \arg(g_{L}^{(n)}g_{R}^{*(n)})$ arises from the numerator, and the loss-induced phase factor $\theta^{(n)}\equiv \arg(\Delta^{(n)}+i\kappa^{(n)}/2)$ originates from the complex denominator part. Importantly, the coherent phase $\phi^{(n)}$ carries the opposite signs between leftward and rightward propagation directions. A virtual excitation transition $Q_L \rightarrow \hat{c}^{(n)} \rightarrow Q_R$ accumulates $+\phi^{(n)}$, while the reversed process accumulates $-\phi^{(n)}$. In contrast, the loss-induced phase $\theta^{(n)}$ remains unchanged, since loss affects both propagation direction identically. This loss-induced phase also renders the effective Hamiltonian non-Hermitian, allowing to break the time-reversal symmetry. For the case with only one lossy channel, these phases make no difference on the magnitudes of leftward and rightward effective couplings, thereby the interaction remains reciprocal. However, when two lossy channels mediate the qubits, the situation has changed. Since $\phi^{(n)}$ reverses signs whereas the loss-induced phase $\theta^{(n)}$ remains invariant, the relative phase between two channels differs in the two propagation directions. The channel can experience constructive interfere in one direction and destructive interfere in the opposite one. This multichannel interference leads to unequal effective coupling strengths, $\left|h_{\leftarrow}\right| \neq \left|h_{\rightarrow}\right|$, enabling nonreciprocal effective interactions between two qubits, as shown in Fig.~\ref{fig1:model}(c).

With some assumptions and transformations, the effective coupling coefficients $h_{\leftarrow}$ and $h_{\rightarrow}$ can also be rewritten as:
\begin{equation}
	h_{\rightleftharpoons}=2Ge^{\mp i\phi_{0}-i\theta_{0}}\cos\left(\frac{\Delta\phi\pm\Delta\theta}{2}\right)
	\label{h_lr}
\end{equation}
Here we assumed equal amplitudes for each channel, i.e., $G^{(1)}=G^{(2)}=G$, which allows to achieve the complete nonreciprocity between two qubits. The following transformations are introduced for simplicity: $\Delta\phi=\phi^{(2)}-\phi^{(1)}$ denotes the relative phases between coherent couplings of two channels, $\Delta\theta=\theta^{(2)}-\theta^{(1)}$ are for the relative loss-induced phases, $\phi_{0}=(\phi^{(1)}+\phi^{(2)})/2$ and $\theta_{0}=(\theta^{(1)}+\theta^{(2)})/2$ are for average coherent coupling phase and loss-induced phase, respectively. Obviously, the effective coupling coefficients in Eq.~(\ref{h_lr}) depend on both $\Delta\phi$ and $\Delta\theta$ simultaneously.

\begin{figure}
	\centering
	\includegraphics{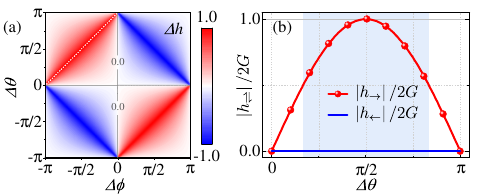}
	\caption{(a) Contour maps of the isolation factor $\Delta h$ versus $\Delta \phi$ and $\Delta \theta$. (b) Normalized rightward or leftward effective coupling strength $\left|h_{\rightleftharpoons}\right|/2G$ as a function of $\Delta \theta$. The case described in (b) is the same as the white dashed curve depicted in (a) with $\Delta \phi=\Delta \theta-\pi$.}
	\label{fig:nonreciprocity}
\end{figure}

To quantify the degree of isolation for effective couplings in different directions, we introduce the isolation factor $\Delta h$, defined in terms of the effective coupling strengths in opposite directions:
\begin{equation}
	\Delta h=\frac{\left|h_\rightarrow\right|-\left|h_\leftarrow\right|}{\left|h_\rightarrow\right|+\left|h_\leftarrow\right|}
\end{equation}
The isolation factor $\Delta h$ ranges from $-1$ to $1$, providing a quantitative measure of isolation. We also plotted $\Delta h$ as a function of both $\Delta\phi$ and $\Delta\theta$ in Fig.~\ref{fig:nonreciprocity}(a), showing the behaviors arising from the interference of coherent and loss-induced phase contributions. When $\Delta h=0$, the magnitudes of the effective couplings in both leftward and rightward directions are equal, i.e., $\left|h_\rightarrow\right|=\left|h_\leftarrow\right|$, indicating the reciprocal coupling between $Q_{L}$ and $Q_{R}$. This condition is fulfilled when either $\Delta \phi= p \pi$ or $\Delta \theta= q \pi$ ($p,q\in \mathbb{Z}$), as predicted by Eq.~(\ref{h_lr}) and illustrated by the gray lines labeled ``0'' in Fig.~\ref{fig:nonreciprocity}(a). Partial nonreciprocity arises when $0 \textless \left|\Delta h\right|\textless 1$, indicating asymmetry effective couplings. And the complete isolation is achieved when the isolation factor $\Delta h=\pm 1$. For instance, $\Delta h=1$ corresponds to $\left|h_\leftarrow\right|=0$, meaning that the effective coupling from $Q_{R}$ to $Q_{L}$ is forbidden and the unidirectional effective coupling from $Q_{L}$ to $Q_{R}$ can be achieved. This isolation condition is in accordance with the result derived from Eq.~(\ref{h_lr}) that $\Delta \phi-\Delta \theta=(2k+1)\pi$ with $\Delta \phi\neq p \pi,\Delta \theta\neq q \pi$ ($p,q\in \mathbb{Z}$) and defines unidirectionality. As a representative case of complete isolation, the condition $\Delta h=1$ and $\Delta \phi=\Delta \theta-\pi$ is highlighted by the white dashed curve in Fig.~\ref{fig:nonreciprocity}(a), which corresponds to the parameter used in Fig.~\ref{fig:nonreciprocity}(b). It is important to note that satisfying $\Delta h=1$ only guarantees the suppression of leftward effective coupling, the rightward effective coupling depends on the loss-induced phase difference $\Delta \theta$, following $\left|h_\rightarrow\right|=2G\left|\sin \Delta \theta\right|$. In particular, the maximum value rightward effective coupling $\left|h_\rightarrow\right|_{\text{max}}=2G$ occurs at $\Delta \theta=\pi/2$ and $\Delta \phi=-\pi /2$, as depicted by the red curve in Fig.~\ref{fig:nonreciprocity}(b). This point also corresponds to the peak value of the normalized coupling strength $\left|h_{\rightleftharpoons}\right|/2G$. And a wide parameter regime exhibiting remarkable nonreciprocity is observed, as shown in the blue shaded region of Fig.~\ref{fig:nonreciprocity}(b).

To explore the transition between nonreciprocal and reciprocal coupling regimes, we analyze the role of both the coherent and loss-induced phase differences. The loss-induced phase $\theta^{(n)}$ is determined solely by the ratio $\Delta^{(n)}/\kappa^{(n)}$ and can thus be engineered the via appropriate design of detunings and decay rates. The phase difference $\Delta \theta$ between two connecting channels is explicitly given by:
\begin{equation}
	\tan \Delta \theta = \frac{2 \left( \Delta^{(1)} \kappa^{(2)} - \Delta^{(2)} \kappa^{(1)} \right)}{4 \Delta^{(1)} \Delta^{(2)} + \kappa^{(1)} \kappa^{(2)}}
	\label{tan-phi}
\end{equation}
Reciprocal coupling occurs when $\Delta \theta =q\pi$ for integer $q$, corresponding to $\tan \Delta \theta =0$. This condition is satisfied when the numerator part of the expression in Eq.~(\ref{tan-phi}) vanishes, i.e., $\Delta^{(1)} \kappa^{(2)} = \Delta^{(2)} \kappa^{(1)}$. In this case, the loss-induced phases in both channels are equal and the effective coupling becomes symmetric. In contrast, unidirectional coupling is achieved when $\Delta \phi-\Delta \theta =(2k+1)\pi$ is satisfied, provided $\Delta \phi\neq p \pi,\Delta \theta\neq q \pi$, breaking the symmetry. However, the maximum unidirectional coupling among them arises only when the condition of $\Delta \theta=\pi/2$ and $\Delta \phi=-\pi /2$ can be simultaneously satisfied. This point corresponds to the optimal constructive interference for rightward coupling and complete destructive interference for leftward coupling, as discussed above.

In our analysis, we fix $\Delta^{(1)} \Delta^{(2)}/\kappa^{(1)} \kappa^{(2)}=-1/4$ to maintain $\Delta \theta=\pi/2$, and tune the coherence phase difference $\Delta \phi$ to achieve the transition between nonreciprocal and reciprocal regimes. Since the loss-induced phase governed by the fixed ratio $\Delta^{(n)}/\kappa^{(n)}$, is typically constrained by the device design and less tunable in situ. The approach to tune the coherent phase is readily achievable in superconducting circuit platforms by precisely tuned qubit-resonator coupling via external microwave drives~\cite{Zeytinoifmmodegelsegfilu2015,Zhou2021}, making our scheme more suitable to engineer the nonreciprocity in practice.\textcolor[rgb]{0,0,0}{The engineered couplings $g_{L/R}^{(n)}$ are obtained by parametrically activated sideband couplings generated by flux modulation~\cite{Strand2013,Naik2017,Roth2017,Caldwell2018,Ma2025}. Therefore, complex-valued couplings with tunbale phases can be engineered in a standard parametric modulation architecture.} A detailed derivation for the dynamically modulation of the qubit-connecting mode coupling is provided in Appendix~\ref{sec:appendixA}. \textcolor[rgb]{0,0,0}{And a representative experimentally feasible parameters of the tunable engineered couplings are provided in Appendix~\ref{sec:appendixB}.}

\section{\label{sec:III}Nonreciprocity between two qubits}
\begin{figure*}
	\centering
	\includegraphics{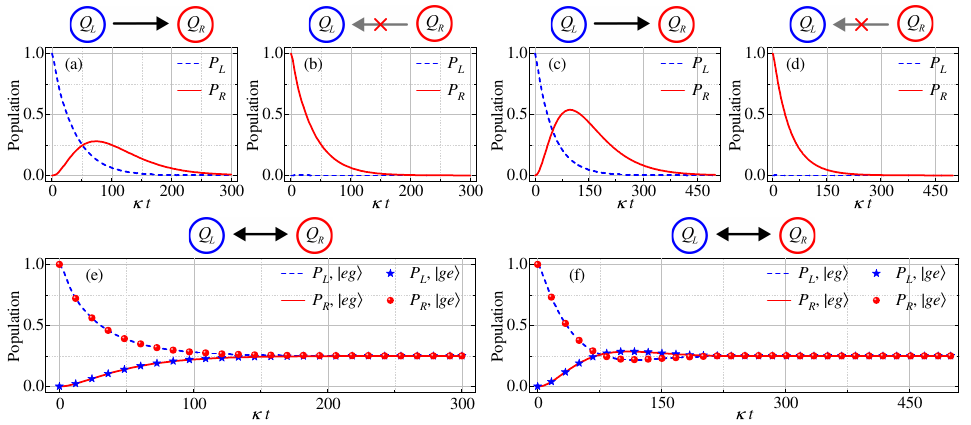}
	\caption{Population dynamics for each qubit under different initial conditions and detuning configurations. (a)-(d) $\Delta \phi=-\pi/2$ for unidirectional coupling case with (a) and (c) the initial state $\left|\psi_0\right\rangle=\left|e\right\rangle_L \left|g\right\rangle_R$ where only $Q_{L}$ is initially excited, while (b) and (d) only $Q_{R}$ initially excited with the initial state $\left|\psi_0\right\rangle=\left|g\right\rangle_L \left|e\right\rangle_R$. (e) and (f) $\Delta \phi=0$ for reciprocal case. $P_L$ and $P_R$ represent the population of $Q_{L}$ and $Q_{R}$, respectively. While the expressions of $\left|eg\right\rangle$ and $\left|ge\right\rangle$ mean the initial states. In the left column [(a), (b) and (e)] $\Delta^{(1)}=-\Delta^{(2)}=\kappa/2$, while in the right column [(c), (d) and (f)] $\Delta^{(1)}=\kappa/200$, $\Delta^{(2)}=-50\kappa$. Other parameters used are  $g^{(1)}_{L}=g^{(1)}_{R}=0.1\sqrt{\kappa\left|\Delta^{(1)}+i\kappa/2\right|}$, $g^{(2)}_{L}=0.1\sqrt{\kappa\left|\Delta^{(2)}+i\kappa/2\right|}$, $g^{(2)}_{R}=g^{(2)}_{L}e^{-i\Delta\phi}$.}
	\label{fig:2qubit-popu}
\end{figure*}
The dynamics of the population for each qubit, derived from Eq.~(\ref{master-eq}), is governed by:
\begin{equation}
	\begin{aligned}
		\frac{d\left\langle\hat{\sigma}_{L}^{+}\hat{\sigma}_{L}^{-}\right\rangle}{dt}=& -i\Lambda_{L}\left\langle\hat{\sigma}_{L}^{+}\hat{\sigma}_{L}^{-}\right\rangle+i\Lambda_{L}\left\langle\hat{\sigma}_{L}^{-}\hat{\sigma}_{L}^{+}\right\rangle \\
		&-ih_{\leftarrow}\left\langle\hat{\sigma}_{L}^{+}\hat{\sigma}_{R}^{-}\right\rangle+ih_{\rightarrow}\left\langle\hat{\sigma}_{L}^{-}\hat{\sigma}_{R}^{+}\right\rangle \\
		\frac{d\left\langle\hat{\sigma}_{R}^{+}\hat{\sigma}_{R}^{-}\right\rangle}{dt}=& -i\Lambda_{R}\left\langle\hat{\sigma}_{R}^{+}\hat{\sigma}_{R}^{-}\right\rangle+i\Lambda_{R}\left\langle\hat{\sigma}_{R}^{-}\hat{\sigma}_{R}^{+}\right\rangle \\
		&-ih_{\rightarrow}\left\langle\hat{\sigma}_{R}^{+}\hat{\sigma}_{L}^{-}\right\rangle+ih_{\leftarrow}\left\langle\hat{\sigma}_{R}^{-}\hat{\sigma}_{L}^{+}\right\rangle
	\end{aligned}
\label{Dynamic_population}
\end{equation}
Under the complete isolation condition with $\Delta h=1$, the effective coupling from $Q_{R}$ to $Q_{L}$ vanishes due to $h_{\leftarrow}=0$. This implies that when the two qubits are initialized with $\left|\psi_0\right\rangle=\left|g\right\rangle_L \left|e\right\rangle_R$, the excitation can not be transferred to the $Q_{L}$. In contrast, the effective coupling $h_{\rightarrow}$ remains finite unless the loss-induced phase difference $\Delta \theta = q\pi$, leading to the excitation of the $Q_{L}$ can transfer to $Q_{R}$ when the initial state $\left|\psi_0\right\rangle=\left|e\right\rangle_L \left|g\right\rangle_R$. This asymmetric excitation  transfer directly reflects the nonreciprocal nature of the interaction.

These behaviors are numerically illustrated in Fig.~\ref{fig:2qubit-popu}, where each system is initialized with only one qubit excited. We first consider the case of complete isolation [Figs.~\ref{fig:2qubit-popu}(a) - (d)], characterized by the coherence phase difference $\Delta \phi=-\pi/2$ and the loss-induced phase difference $\Delta \theta=\pi/2$, which corresponds to the condition to achieve maximum unidirectional coupling from $Q_{L}$ to $Q_{R}$. In Figs.~\ref{fig:2qubit-popu}(a) and (c), where only $Q_{L}$ is initially excited, clear population transfer from $Q_{L}$ to $Q_{R}$ is observed, confirming the rightward effective coupling. However, when only $Q_{R}$ is initially excited, as shown in Figs.~\ref{fig:2qubit-popu}(b) and (d), the excitation remains localized in $Q_{R}$, indicating complete suppression of the leftward coupling. As a reference, the reciprocal case with $\Delta \phi=0$ is shown in Figs.~\ref{fig:2qubit-popu}(e) and (f), where identical leftward and rightward couplings lead to symmetric population oscillations, regardless of the initially excited qubit.

A key observation lies in the difference in oscillation amplitude between the left column [Figs.~\ref{fig:2qubit-popu}(a), (b) and (e)] and the right column [Figs.~\ref{fig:2qubit-popu}(c), (d) and (f)]. Despite both detuning configurations satisfying maximum unidirectional effective coupling strength $\left|h_\rightarrow\right|_{\text{max}}=2G$, the excitation received by $Q_{R}$ differs, as depicted by the red curves in Figs.~\ref{fig:2qubit-popu}(a) and (c). Specifically, $Q_{R}$ exhibits a larger population amplitude in Fig.~\ref{fig:2qubit-popu}(c) than in (a), showing enhanced excitation transfer under the corresponding configuration. This difference originates from the distinct response of the connecting modes rather than from the effective coupling strength alone. While the effective coupling strength determines the directional interaction in the adiabatic limit, the population dynamics are governed by the full response of the lossy connecting modes. When both connecting modes are detuned $\Delta^{(1)}=-\Delta^{(2)}=\kappa/2$, the qubit excitation is distributed over lossy channels, leading to a reduced population amplitude. In contrast, for strongly asymmetric detunings with $\Delta^{(1)}=\kappa/200$, $\Delta^{(2)}=-50\kappa$, the excitation transfer is dominated by a near-resonant channel, while the far-detuned channel primarily contributes a phase shift. This asymmetric response enhance the excitation transfer amplitude, providing an additional degree of freedom for optimizing nonreciprocal system design.

\section{\label{sec:IV}Nonreciprocal entanglement between two qubits}
\begin{figure}
	\centering
	\includegraphics{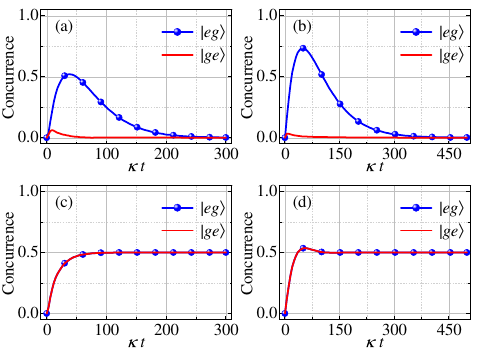}
	\caption{Time evolution of concurrence between two qubits under various condition: (a) and (b) nonreciprocal transient entanglement for phase $\Delta \phi=-\pi/2$, (c) and (d) reciprocal entanglement with $\Delta \phi=0$. The legend of $\left|eg\right\rangle$ and $\left|ge\right\rangle$ denote the dynamics of concurrence with the initial state $\left|\psi_0\right\rangle=\left|e\right\rangle_L \left|g\right\rangle_R$ and $\left|\psi_0\right\rangle=\left|g\right\rangle_L \left|e\right\rangle_R$, respectively. The parameters used are $\Delta^{(1)}=-\Delta^{(2)}=\kappa/2$ in (a) and (c), while $\Delta^{(1)}=\kappa/200$, $\Delta^{(2)}=-50\kappa$ in (b) and (d).}
	\label{fig:2qubit-con}
\end{figure}
So far, we have discussed the nonreciprocal behavior in population transfer between qubits. In this section, we turn to the entanglement properties between two qubits within our scheme. To quantify the entanglement, we adopt the concurrence $C$, a widely used measure for bipartite qubit systems and originally introduced by Wootters~\cite{Wootters1998}. The explicit formula of the concurrence, for a pair of qubits with a given density matrix $\rho$, is defined as: $C(\rho) = \max \left\{0,\lambda_1-\lambda_2-\lambda_3-\lambda_4\right\}$. Here the $\lambda_i$s are the square roots of the eigenvalues of the non-Hermitian matrix $\rho \tilde{\rho}$, arranged in decreasing order. The matrix $\tilde{\rho}$ is the spin-flipped counterpart of $\rho$, defined by $\tilde{\rho} = \left( \sigma_{y} \otimes \sigma_{y} \right) \rho^{*} \left( \sigma_{y} \otimes \sigma_{y} \right)$ with $\rho^{*}$ the complex conjugate of $\rho$ in the standard basis.

The time evolution of concurrence is plotted in Fig.~\ref{fig:2qubit-con}, where the system is initially prepared with only one qubit excited. Under the complete isolation condition with $\Delta h= 1$ and the maximum effective coupling with $\Delta \theta=\pi/2$ and $\Delta \phi=-\pi/2$, the dynamics of concurrence exhibit different behaviors depending on which qubit is initially excited. Specifically, when the initial state is $\left|\psi_0\right\rangle=\left|e\right\rangle_L \left|g\right\rangle_R$, the concurrence occurs an notable growth over time [blue curves in Figs.~\ref{fig:2qubit-con}(a) and (b)], indicating the generation of entanglement between two qubits due to rightward excitation transfer. In contrast, when the system is initialized with only $Q_{R}$ excited, the concurrence remains nearly 0 throughout the entire evolution, as depicted with red curves in Figs.~\ref{fig:2qubit-con}(a) and (b). This vanish originates from the complete suppression of the leftward effective coupling. For comparison, in the reciprocal case with $\Delta h= 0$ and $\Delta \phi=0$, the concurrence have the identical evolution under the exchange of excited qubits, as shown in Figs.~\ref{fig:2qubit-con} (c) and (d). The blue and red curves coincide perfectly, showing the reciprocity of entanglement dynamics in this configuration.

A notable feature emerges when comparing the left column [Figs.~\ref{fig:2qubit-con} (a) and (c)] and the right column [Figs.~\ref{fig:2qubit-con} (b) and (d)]. The maximum value of $C$ in Fig.~\ref{fig:2qubit-con} (b) is much larger than that in Fig.~\ref{fig:2qubit-con} (a), implying a stronger degree of entanglement under the detuning configuration $\Delta^{(1)}=\kappa/200$, $\Delta^{(2)}=-50\kappa$, compared to the symmetric configuration. This behavior is consistent with the corresponding population transfer previously discussed. Moreover, in the reciprocal case with these two sets of parameters, both detuning configurations perform nearly identical entanglement evolution, with the concurrence approaching a steady value of 0.5, suggesting a robust entanglement generation under reciprocal coupling.

\section{\label{sec:V}The influence of decoherence on nonreciprocity}
In the previous section, we discussed the system neglecting the decoherence of qubits including both qubit dephasing and energy relaxation. Now we consider the influence of decoherence. In this case, the master equation is modified by:
\begin{equation}
	\begin{split}
		\frac{d\hat{\rho}}{dt}&=-i\left[\hat{H}^{\prime},\hat{\rho}\right]+\sum_{n=1}^{2}\kappa^{(n)}\mathcal{D}\left[\hat{c}^{(n)}\right]\hat{\rho} \\
		&+\sum_{m=L,R}\gamma_{m}\mathcal{D}\left[\hat{\sigma}_{m}^{-}\right]\hat{\rho}+\sum_{m=L,R}\frac{\gamma_{\varphi m}}{2}\mathcal{D}\left[\hat{\sigma}_{m}^{z}\right]\hat{\rho}
	\end{split}
	\label{master-eq+decoherence}
\end{equation}
where $\gamma_{m}$ and $\gamma_{\varphi m}$ are the energy relaxation rate and pure dephasing rate of qubit $Q_{m}$, respectively. For simplicity, we take $\gamma_{L}=\gamma_{R}=\gamma$ and $\gamma_{\varphi L}=\gamma_{\varphi R}=\gamma_{\varphi}$. The dynamics of qubit populations is:
\begin{equation}
	\begin{aligned}
		\frac{d\left\langle\hat{\sigma}_{L}^{+}\hat{\sigma}_{L}^{-}\right\rangle}{dt}=& -\left(i\Lambda_{L}+\gamma\right)\left\langle\hat{\sigma}_{L}^{+}\hat{\sigma}_{L}^{-}\right\rangle+i\Lambda_{L}\left\langle\hat{\sigma}_{L}^{-}\hat{\sigma}_{L}^{+}\right\rangle \\
		&-ih_{\leftarrow}\left\langle\hat{\sigma}_{L}^{+}\hat{\sigma}_{R}^{-}\right\rangle+ih_{\rightarrow}\left\langle\hat{\sigma}_{L}^{-}\hat{\sigma}_{R}^{+}\right\rangle \\
		\frac{d\left\langle\hat{\sigma}_{R}^{+}\hat{\sigma}_{R}^{-}\right\rangle}{dt}=& -\left(i\Lambda_{R}+\gamma\right)\left\langle\hat{\sigma}_{R}^{+}\hat{\sigma}_{R}^{-}\right\rangle+i\Lambda_{R}\left\langle\hat{\sigma}_{R}^{-}\hat{\sigma}_{R}^{+}\right\rangle \\
		&-ih_{\rightarrow}\left\langle\hat{\sigma}_{R}^{+}\hat{\sigma}_{L}^{-}\right\rangle+ih_{\leftarrow}\left\langle\hat{\sigma}_{R}^{-}\hat{\sigma}_{L}^{+}\right\rangle
	\end{aligned}
	\label{Dynamic_population+decoherence}
\end{equation}
\begin{figure}
	\centering
	\includegraphics{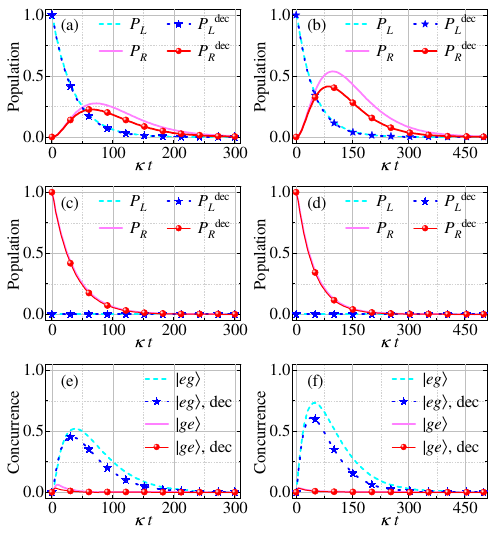}
	\caption{Dynamics of qubit populations and concurrence with and without decoherence, in complete isolation condition. (a)-(d) Time evolution of qubit populations: (a) and (b) initialized with only $Q_{L}$ excited, while (c) and (d) initialized with only $Q_{R}$ excited. $P_i^{~\text{dec}}$ represent the case consider decoherence including qubit energy relaxation $\gamma=10^{-3}\kappa$ and qubit dephasing $\gamma_{\varphi}=3\times10^{-3}\kappa$. While $P_L$ and $P_R$ represent the population of $Q_{L}$ and $Q_{R}$ without decoherence. (e)-(f) Time evolution of concurrence $C$ with the initial state $\left|eg\right\rangle$ and $\left|ge\right\rangle$. The texts with and without $\text{dec}$ in (e)-(f) correspond to the case with and without decoherence, respectively. Other parameters used are $\Delta^{(1)}=-\Delta^{(2)}=\kappa/2$ for the left column [(a), (c), and (e)], while $\Delta^{(1)}=\kappa/200$, $\Delta^{(2)}=-50\kappa$ for the right column [(b) ,(d) and (f)].}
	\label{fig:decoherence}
\end{figure}

Physically, additional qubit dephasing and energy relaxation processes can be incorporated to Eq.~(\ref{master-eq+decoherence}) as local Lindblad terms, leading to a finite lifetime of excitations and a suppression of population amplitude. However, qubit decoherence does not affect the structure of the effective coupling coefficient that describe the nonreciprocal response, as evident from Eq.~(\ref{Dynamic_population+decoherence}). Consequently, the isolation condition determined solely by effective coupling coefficient will not be affected in the presence of decoherence.

To quantitatively illustrate the influence of qubit decoherence, we perform a direct comparison between the dynamics with and without qubit decoherence. Figs.~\ref{fig:decoherence}(a-d) show the evolution of qubit populations $P_i$ in both detuning configurations, where $P_i^{~\text{dec}}$ ($P_i$) denotes the case with (without) qubit decoherence. Similarly, Figs.~\ref{fig:decoherence}(e-f) present the concurrence dynamics for different initial states, again comparing the cases with and without decoherence in both detuning configurations. The results show that qubit decoherence has only a limited influence on the amplitude of qubit populations and concurrence. However, both the nonreciprocal excitation transfer and nonreciprocal entanglement remain evident even in the presence of decoherence, highlighting the feasibility of implementing the proposed nonreciprocal coupling in realistic superconducting quantum circuits. From the perspective of quantum information processing, this robustness of nonreciprocity under realistic decoherence is important for potential applications in isolating quantum channels.

\section{\label{sec:VI}Influence of qubit frequency detuning on nonreciprocity}
\begin{figure*}
	\centering
	\includegraphics{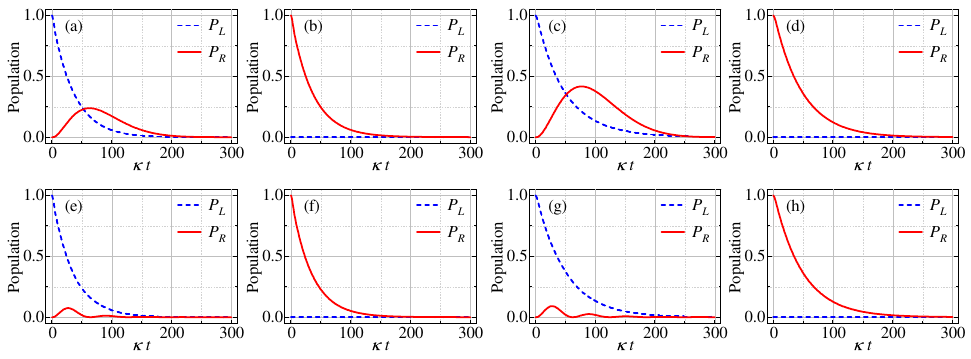}[h]
	\caption{Dynamics of qubit populations with different qubit detunings in complete isolation condition. (a)-(d) Time evolution of qubit populations with $\Delta_R=\kappa/50$: (a) and (c) initialized with only $Q_{L}$ excited, while (b) and (d) initialized with only $Q_{R}$ excited. (e)-(f) Time evolution of of qubit populations with $\Delta_R=\kappa/10$: (e) and (g) initialized with only $Q_{L}$ excited, while (f) and (h) initialized with only $Q_{R}$ excited. Other parameters used are $\Delta^{(1)}=-\Delta^{(2)}=\kappa/2$ for the left column [(a), (b), (e) and (f)], while $\Delta^{(1)}=\kappa/200$, $\Delta^{(2)}=-50\kappa$ for the right column [(c) ,(d) ,(g) and (h)].}
	\label{fig:detuning}
\end{figure*}
The analysis presented above was carried out under the assumption of resonant qubits, $\omega_L=\omega_R=\omega_0$, which simplifies the analytical treatment. We now lift this restriction and analyze the impact of a finite qubit frequency mismatch on the nonreciprocal dynamics.

We consider the general case where the qubits are not exactly resonant $\omega_L\neq\omega_R$. In the rotating frame defined by the reference frequency $\omega_0$, the total Hamiltonian in Eq.~(\ref{H_tot}) is transformed as:
\begin{equation}
	\begin{split}
		\hat{H}_{\text{nr}}^\prime = & -\sum_{m=L,R} \frac{1}{2} \Delta_m \hat{\sigma}_m^z 
		- \sum_{n=1}^2 \Delta^{(n)} \hat{c}^{(n)\dagger} \hat{c}^{(n)} \\
		+ & \sum_{n=1}^2 \left[ \left( \lambda_{L}^{(n)} \hat{\sigma}_{L}^+ + \lambda_{R}^{(n)} \hat{\sigma}_{R}^+ \right) \hat{c}^{(n)} + H.c. \right]
	\end{split}
	\label{H_detuning}
\end{equation}
where $\Delta_m=\omega_0-\omega_m$ represents the detuning of $Q_m$ in the rotating frame. Following the same adiabatic elimination procedure described in Sec.~\ref{sec:II} of the main text, the lossy connecting modes can be eliminated. This leads to the effective equations of motion for the qubit lowering operators that keep the same nonreciprocal structure as in the resonant case:
\begin{equation}
	\begin{split}
		\frac{d\hat{\sigma}_{L}^{-}}{dt}=& -i(\Lambda_{L}-\Delta_{L})\hat{\sigma}_{L}^{-}+ ih_{\leftarrow}\hat{\sigma}_{L}^{z}\hat{\sigma}_{R}^{-} \\
		\frac{d\hat{\sigma}_{R}^{-}}{dt}=& -i(\Lambda_{R}-\Delta_{R})\hat{\sigma}_{R}^{-}+ ih_{\rightarrow}\hat{\sigma}_{R}^{z}\hat{\sigma}_{L}^{-}
	\end{split}
	\label{Dynamic_lowering_detuning}
\end{equation}

Importantly, the qubit detunings $\Delta_m$ only enter as local terms and do not modify the effective coupling coefficients $h_{\leftarrow}$ and $h_{\rightarrow}$. As a result, qubit frequency mismatch does not alter the fundamental nonreciprocal character of the dynamics.

This robustness is further illustrated by numerical simulations of the full master equation including finite qubit detunings. For simplicity, we fix $\omega_L=\omega_0$ and vary the detuning of the right qubit $\Delta_R$. As depicted in Fig.~\ref{fig:detuning}, we consider two representative qubit detunings of the $Q_R$, $\Delta_R=\kappa/50$ (corresponding to 1 $\mathrm{MHz}$) and $\Delta_R=\kappa/10$ (corresponding to 5 $\mathrm{MHz}$). In both cases, the excitation transfer remains nonreciprocal. For small qubit detunings ($\Delta_R=\kappa/50$), the primary influence of qubit detuning on nonreciprocity is the amplitude reduction in the excitation received by $Q_R$, while the directionality of excitation transfer is preserved. As the qubit detunings increases ($\Delta_R=\kappa/10$), the dynamics of qubit population exhibit oscillations due to the detuned evolution. Nevertheless, the nonreciprocal behavior persists. Specifically, when $Q_R$ is initially excited, the population of $Q_L$ remains strictly zero, as indicated by the blue curves in Figs.~\ref{fig:detuning} (b), (d), (f) and (h), indicating the robustness of the nonreciprocal dynamics. These results confirm that the proposed loss-induced nonreciprocal scheme is robust against realistic qubit frequency mismatches in a certain range and does not rely on strict qubit resonance. 

\section{\label{sec:VII}Experimental feasibility}
In this section we discuss the experimental feasibility of our scheme. Our proposal can be readily implemented with the superconducting transmon qubits coupled via resonators in the circuit QED architecture. In typical experiments, the device operates in the dilution refrigerator at temperature of about 10~mK, ensuring that qubits maintain well superconductivity and thermal noise can be neglected~\cite{Krinner2019}. As the most widely used qubit currently~\cite{Kjaergaard2020}, transmon qubit strongly suppresses charge noise while maintaining sufficient anharmonicity for selective control and offers relative simplicity of fabrication~\cite{Koch2007}, making it an ideal choice for our scheme. Transmons are routinely fabricated with $Al/AlO_{x}$ Josephson junctions and can be capacitively coupled to coplanar waveguide or 3D resonators for both readout and mediated interactions~\cite{Krantz2019,Blais2021}. To enable frequency tunability, the transmon qubit incorporates a SQUID loop, whose effective Josephson energy can be modulated by an external magnetic flux. The required flux $\Phi_{\text{ext}}(t)$ is applied through a flux bias line, allowing the frequency of each qubit to be tuned rapidly and independently~\cite{Manenti2021,Didier2018}. By utilizing advanced chip design and manufacturing technologies, the coherent times $T_1$ and $T_2$ of transmon qubits can reach millisecond order~\cite{Bland2025,Tuokkola2025,Wang2022}.
 
It has been demonstrated that resonators decay rates $\kappa/2\pi$ can be engineered over a wide range, from a few $\mathrm{kHz}$ to $50 ~\mathrm{MHz}$~\cite{Sunada2022,Potts2025}. The required qubit-resonator detunings are easily achieved by device design or flux tuning, spanning hundreds of $\mathrm{kHz}$ to several $\mathrm{GHz}$~\cite{Blais2021}. These parameters allow one to satisfy the key condition $\Delta^{(1)} \Delta^{(2)}/\kappa^{(1)} \kappa^{(2)}=-1/4$, corresponding to a loss-induced phase difference $\Delta \theta=\pi/2$. Moreover, parametric driving techniques allow in situ tuning of both the amplitude and phase of the coupling effective couplings, enabling control of coherent phase difference $\Delta \phi$~\cite{Zhou2021,Zeytinoifmmodegelsegfilu2015}. 

\section{\label{sec:VIII}Conclusion}
We have proposed a scheme to achieve nonreciprocity and nonreciprocal entanglement between two superconducting qubits mediated by lossy connecting channels. The nonreciprocity arises from the interference between multiple connecting channels, combined with the distinct contributions of coherent and loss-induced phases to the effective coupling strengths in opposite directions. By engineering the loss and detuning of resonators, the effective qubit-qubit coupling induced by the connecting modes can be tuned, enabling the system to reach the complete isolation regime and the unidirectional population transfer. Numerical simulations reveal that the evolution of qubit populations depends not only on the effective coupling strengths but also on the detuning asymmetry between two channels, providing an additional degree for optimizing population transfer amplitude. Extending the analysis to entanglement, we have also show the unidirectional generation of entanglement, for concurrence exhibiting significant asymmetry depending on the initially excited qubit. Moreover, the influence of qubit decoherence including both the qubit dephasing and energy relaxation have also been analyzed, showing the reduction of population while the remaining of nonreciprocal behavior. Our loss-induced scheme to achieve nonreciprocity in superconducting platform does not require magnetic fields, nonlinearity, or spatially adjacent, enabling the controllable nonreciprocity between distant qubits using only engineered loss. This work provides a promising and tunable platform for directional quantum information flow, offering potential applications in scalable quantum networks.

\appendix
\begin{color}[rgb]{0,0,0}
\section{\label{sec:appendixA}The derivation for the dynamically modulation of the qubit-connecting mode coupling}
In this appendix, we provide a derivation of the complex and independently tunable qubit-resonator couplings used in Eq.~(\ref{H_prime}) of the main text. Our goal here is to present a concrete implementation scheme to show how the engineered effective Hamiltonian $\hat{H}^\prime$ in Eq.~(\ref{H_prime}) can be generated by parametric modulation. To keep the derivation fully consistent with the frame adopted throughout the paper, we first transform the laboratory-frame Hamiltonian to the global rotating frame at average qubit frequency $\omega_0$, then do some rotating frame transformation to remove the free terms, finally apply the rotating-wave approximation to obtain the engineered Hamiltonian. This procedure yields the complex engineered couplings in the effective slow timescale Hamiltonian, which governs the same dynamics in the global rotating frame as the Hamiltonian presented in the main text. For clarity, we first consider a generic system consisting of a qubit coupled to two connecting modes $\hat{c}^{(1)}$ and $\hat{c}^{(2)}$. The generalization to two qubits is given at the end of this appendix.

\subsection{The setup in the laboratory frame}	
In the laboratory frame, the Hamiltonian of the system can be written as $(\hbar=1)$:
\begin{equation}
	\hat{H}_{\text{lab}} = \hat{H}_q(t)+\hat{H}_c + \hat{H}_{\text{int}}(t).
\end{equation}
Here, the qubit Hamiltonian is $\hat{H}_q(t)=\frac{1}{2}\omega_m(t)\hat{\sigma}^{z}$, where $\omega_m(t)$ is time-dependent due to flux modulation. The connecting modes are described by: $\hat{H}_c=\sum_{n=1}^2 \omega_c^{(n)}\hat{c}^{(n)\dagger} \hat{c}^{(n)}$ with $\omega_c^{(n)}$ the frequencies of the connecting modes. The interaction between the qubit and the connecting modes, typically originating from either capacitive or inductive coupling, takes the form:
\begin{equation}
	\hat{H}_{\text{int}}(t) = \sum_{n=1}^2 \lambda^{(n)} (\hat{\sigma}^+\hat{c}^{(n)} +   \hat{\sigma}^-\hat{c}^{(n)\dagger}).
	\label{H_int(t)}
\end{equation}
where $\lambda^{(n)}$ denotes the real bare coupling strength between the qubit and the $n$-th connecting mode $\hat{c}^{(n)}$. 

To simultaneously activate two distinct interactions with $\hat{c}^{(1)}$ and $\hat{c}^{(2)}$, we apply a two-tone flux modulation to the qubit~\cite{Zhou2021}. The frequency of qubit is thus composed of a static component and two independently controlled AC components~\cite{Strand2013,Naik2017,Roth2017,Caldwell2018,Ma2025}:
\begin{equation}
	\omega_m(t)=\omega_0+A_1\cos(\omega_{d1}t+\psi_1)+A_2\cos(\omega_{d2}t+\psi_2)
\end{equation}
where $\omega_0$ is the average qubit frequency. The modulation parameters ($A_1$, $\omega_{d1}$, $\psi_1$) and ($A_2$, $\omega_{d2}$, $\psi_2$) are independent and externally controllable, enabling selective activation of the required sideband in each coupling channel.

\subsection{Rotating frame at $\omega_0$}
We now move to the same global rotating frame used in the main text, defined by the average qubit frequency $\omega_0$ via the unitary operator:
\begin{equation}
	\hat{U}_{\text{rot}}(t) = \exp\left[-i\omega_0\left(\frac{1}{2}\hat{\sigma}^z+\sum_{n=1}^{2} \hat{c}^{(n)\dagger} \hat{c}^{(n)}\right) t\right].
\end{equation}
The laboratory Hamiltonian transforms as $\hat{H}_{\text{rot}}(t) = \hat{U}^{\dagger}_{\text{rot}}(t)\hat{H}_{\text{lab}}(t)\hat{U}_{\text{rot}}(t)-i\hat{U}^{\dagger}_{\text{rot}}(t) \dot{\hat{U}}_{\text{rot}} $, takes the form:
\begin{equation}
	\begin{aligned}
		\hat{H}_{\text{rot}}(t) =& \frac{1}{2}\delta\omega(t)\hat{\sigma}^z-\sum_{n=1}^2\Delta_0^{(n)}\hat{c}^{(n)\dagger} \hat{c}^{(n)} \\
		+&\sum_{n=1}^2 \lambda^{(n)}\left[ \hat{\sigma}^+ \hat{c}^{(n)} + \text{H.c.} \right]
	\end{aligned}
	\label{H_rot}
\end{equation}
where $\delta\omega(t)=\omega_m(t)-\omega_0=A_1\cos(\omega_{d1}t+\psi_1)+A_2\cos(\omega_{d2}t+\psi_2)$ and $\Delta_0^{(n)}=\omega_0-\omega_c^{(n)}$ is the bare detuning of the $n$-th connecting mode from the rotating frame frequency $\omega_0$. From Eq.~(\ref{H_rot}), we can find that the qubit-resonator coupling keeps the Jaynes-Cummings form, while the mode frequencies perform as detunings and the qubit frequency modulation remains as a time-dependent local qubit term.

\subsection{Removal of the qubit modulation term}
To expose the sideband structure induced by the qubit-frequency modulation, we remove the local qubit term with respect to the Hamiltonian $\frac{1}{2}\delta\omega(t)\hat{\sigma}^z$. The transformation is given by the unitary operator $\hat{U}_{m}(t) = \mathcal{T}\exp\left[-i\int_0^t \delta\omega(\tau) d\tau\right]$. Since $\left[\delta\omega(t_1)\hat{\sigma}^z, \delta\omega(t_2)\hat{\sigma}^z\right] = 0$ for $\forall t_1, t_2 $, the time-ordering operator $\mathcal{T}$ can be dropped:
\begin{equation}
	\hat{U}_{m}(t) = \exp\left[-i\Phi(t)\frac{\hat{\sigma}^z}{2} \right].
\end{equation}
where $\Phi(t)=\int_0^t \delta\omega(\tau) d\tau$. Transforming $\hat{H}_{\text{rot}}(t)$ to this rotating frame, the Hamiltonian reads $\hat{H}_{m}(t)=\hat{U}_{m}^{\dagger}\hat{H}_{\text{rot}}\hat{U}_{m}-i\hat{U}_{m}^{\dagger}\dot{\hat{U}}_{m}$. Using $\hat{U}_{m}^{\dagger}\sigma^+\hat{U}_{m}=\sigma^+e^{i\Phi(t)}$, we can find that the explicit modulation term of qubit has been absorbed into the time-dependent phase factors $e^{i\Phi(t)}$ of the coupling terms:
\begin{equation}
	\hat{H}_{m}(t)=-\sum_{n=1}^2\Delta_0^{(n)}\hat{c}^{(n)\dagger} \hat{c}^{(n)}+\sum_{n=1}^{2}\lambda^{(n)} \left[\hat{\sigma}^+\hat{c}^{(n)}e^{i\Phi(t)} + \text{H.c.}\right]
	\label{H_m}
\end{equation}

\subsection{Interaction picture with respect to the bare detuning}
Then we introduce an interaction picture associate with the bare connecting mode detuning:
\begin{equation}
	\hat{U}_{\Delta_0}(t)=\exp\left[i\sum_{n=1}^2\Delta_0^{(n)}\hat{c}^{(n)\dagger}\hat{c}^{(n)}t\right].
	\label{Udelta0}
\end{equation}
And the corresponding Hamiltonian is transformed as:
\begin{equation}
	\hat{H}_{I}(t)=\sum_{n=1}^{2}\lambda^{(n)} \left[\hat{\sigma}^+\hat{c}^{(n)}e^{i\Phi(t)}e^{i\Delta_0^{(n)} t} + \text{H.c.}\right]
	\label{H_I}
\end{equation}
For simplicity, we define the phase exponent for the $n$-th connecting mode as $\alpha^{(n)}(t) \equiv  \Phi(t) + \Delta_0^{(n)}t=\int_0^t \delta\omega(\tau) d\tau + \Delta_0^{(n)}t$. Thus, the interaction Hamiltonian in Eq.~(\ref{H_I}) can be rewritten as: $\hat{H}_{I}(t)=\sum_{n=1}^{2}\lambda^{(n)} \left[\hat{\sigma}^+\hat{c}^{(n)}e^{i\alpha^{(n)}t} + \text{H.c.}\right]$.

\subsection{Jacobi-Anger expansion}
Integrating the modulation phase and substituting it into $\alpha^{(n)}(t)$, we obtain:
\begin{equation}
	\begin{aligned}
		\alpha^{(n)}(t)=&\Delta_0^{(n)}t+ \frac{A_1}{\omega_{d1}}\sin(\omega_{d1}t + \psi_1) \\
		& + \frac{A_2}{\omega_{d2}}\sin(\omega_{d2}t + \psi_2) + \mathrm{const.},
	\end{aligned}
\end{equation}
where the constant term only contributes a global phase and can be omitted.
	
To proceed, we apply the Jacobi-Anger expansion $e^{iz\sin\Theta}=\sum_k J_k(z)e^{ik\Theta}$, where $J_k(z)$ is the $k$-th order Bessel function of the first kind, to expand the exponential term $e^{i\alpha^{(n)}(t)}$ for both drive tones:
\begin{equation}
	\begin{aligned}
		e^{i\alpha^{(n)}(t)} = e^{i\Delta_0^{(n)}t} &\cdot \left[ \sum_{k=-\infty}^{\infty} J_k\left(\frac{A_1}{\omega_{d1}}\right) e^{ik(\omega_{d1}t + \psi_1)} \right] \\
		&\cdot \left[ \sum_{l=-\infty}^{\infty} J_l\left(\frac{A_2}{\omega_{d2}}\right) e^{il(\omega_{d2}t + \psi_2)} \right].
	\end{aligned} 
\end{equation}
The interaction Hamiltonian $\hat{H}_{I}(t)$ becomes a double summation over all possible frequency sidebands $k$ and $l$:
\begin{equation}
	\begin{aligned}
		\hat{H}_I(t)=&\sum_{n=1}^2 \lambda^{(n)}\left[\hat{\sigma}^+ \hat{c}^{(n)} \sum_{k,l=-\infty}^{\infty} J_k\left(\frac{A_1}{\omega_{d1}}\right) J_l\left(\frac{A_2}{\omega_{d2}}\right) \cdot \right. \\
		& \left. e^{i\left(\Delta_0^{(n)}+k\omega_{d1}+l\omega_{d2} \right)t} e^{i(k\psi_1+l\psi_2)} + \text{H.c.} \right].
	\end{aligned}
	\label{H_possible}
\end{equation}
This equation shows that the interaction contains an infinite set of modulation-induced sidebands oscillating at frequencies $\Delta_0^{(n)}+k\omega_{d1}+l\omega_{d2}$, where $k,l\in \mathbb{Z}$ are the sideband orders associated with the two-tone modulation. 
	
\subsection{Sideband selection with residual engineered detunings}
To activate the desired couplings to the two connecting modes while retaining residual engineered detunings, the drive frequencies are chosen to compensate the bare detunings up to residual detunings. Specifically, for each connecting mode we impose the sideband condition $\Delta_0^{(n)}+k\omega_{d1}+l\omega_{d2}=\Delta^{(n)}$ with $\Delta^{(n)}$ the residual engineered detuning that remains after sideband selection. For the coupling between the qubit and the mode $\hat{c}^{(1)}$, we select the $k=\pm1$, $l=0$ sideband by setting the first drive frequency to: $\Delta_0^{(1)}\pm\omega_{d1}=\Delta^{(1)}$, where the sign is determined by choosing positive drive frequencies $\omega_{d1},\omega_{d2} > 0$ together with the appropriate sideband orders. Similarly, to simultaneously activate the coupling to $\hat{c}^{(2)}$, we apply the second drive tone and choose the $k=0$, $l=\pm1$ sideband for the $n=2$ channel by setting: $\Delta_0^{(2)}\pm\omega_{d2}=\Delta^{(2)}$. Under these conditions, the corresponding selected sideband contributions oscillate only at the residual frequencies $\Delta^{(n)}$, while all other sidebands in Eq.~(\ref{H_possible}) remain rapidly oscillating.
	
Keeping only these selected contributions and neglecting all rapidly oscillating terms within the rotating-wave approximation, the resulting slow-timescale Hamiltonian is:
\begin{equation}
	\hat{H}_{\text{slow}}=\sum_{n=1}^2 \left[g^{(n)}e^{i\Delta^{(n)}}\hat{\sigma}^+ \hat{c}^{(n)} + \text{H.c.} \right],
	\label{H_slow}
\end{equation}
with the complex engineered coupling coefficients:
\begin{equation}
	\begin{aligned}
		g^{(1)}&=\lambda^{(1)} J_1\left(\frac{A_1}{\omega_{d1}}\right) J_0\left(\frac{A_2}{\omega_{d2}}\right) e^{i\tilde{\psi}_1}  \\
		g^{(2)}&=\lambda^{(2)} J_0\left(\frac{A_1}{\omega_{d1}}\right) J_1\left(\frac{A_2}{\omega_{d2}}\right) e^{i\tilde{\psi}_2}.
	\end{aligned}
	\label{g^n}
\end{equation}
Here, the RWA is valid provided that the selected resonant sidebands are well isolated in frequency space, which means all other non-selected sidebands satisfy $\left|\Delta_0^{(n)}+k\omega_{d1}+l\omega_{d2}-\Delta^{(n)}\right| \gg \left|\lambda^{(n)} J_k\left(\frac{A_1}{\omega_{d1}}\right) J_l\left(\frac{A_2}{\omega_{d2}}\right)\right|$. Under this condition, the non-selected terms average out on the timescale relevant to the engineered slow dynamics, and only the selected sidebands contribute effectively. 
	
Note that we modify the phase of the engineered complex coupling as $\tilde{\psi}_1$ and $\tilde{\psi}_2$, including both the modulation phases and additional $\pi$-shift arising from the selected sideband orders. In particular, when a negative order sideband is selected, $J_{-1}(x)=-J_{1}(x)$ introduce an additional minus sign, which is absorbed into the corresponding phase of the engineered complex coupling. Therefore, Eq.~(\ref{g^n}) show that the magnitudes of the engineered couplings are controlled by the modulation amplitudes via the Bessel functions ($A_1$ and $A_2$), while their phases are originated from the modulation phases ($e^{i\psi_1}$ and $e^{i\psi_2}$) together with the sideband order. Thus, the qubit-connecting mode couplings are independently tunable in both magnitudes and phases.
	
\subsection{Transformation to the residual engineered detuning frame}
Next, we absorb the phase factors $e^{i\Delta^{(n)}t}$ in Eq.~(\ref{H_slow}) by introducing the residual engineered detuning frame:
\begin{equation}
	\hat{U}_{\text{res}}(t)=\exp\left[-i\sum_{n=1}^2\Delta^{(n)}\hat{c}^{(n)\dagger}\hat{c}^{(n)}t\right].
	\label{U_res}
\end{equation}
Using $\hat{U}_{\text{res}}^{\dagger}\hat{c}^{(n)}\hat{U}_{\text{res}}=\hat{c}^{(n)}e^{-i\Delta^{(n)}t}$, the slow Hamiltonian $\hat{H}_{\text{slow}}$ is transformed as $\hat{H}_{\text{eng}}=\hat{U}_{\text{res}}^{\dagger}\hat{H}_{\text{slow}}\hat{U}_{\text{res}}-i\hat{U}_{\text{res}}^{\dagger}\dot{\hat{U}}_{\text{res}}$. Therefore, we obtain the static engineered effective form:
\begin{equation}
	\hat{H}_{\text{eng}}=-\sum_{n=1}^2\Delta^{(n)}\hat{c}^{(n)\dagger}\hat{c}^{(n)}+\sum_{n=1}^2 \left[g^{(n)}\hat{\sigma}^+ \hat{c}^{(n)} + \text{H.c.} \right],
	\label{H_eng}
\end{equation}
This is the desired effective Hamiltonian in slow timescale with engineered complex couplings and residual engineered detunings.

\begin{figure}
	\centering
	\includegraphics{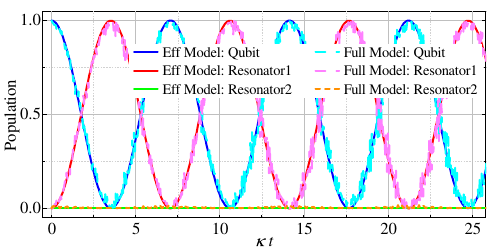}
	\caption{\textcolor[rgb]{0,0,0}{Numerical simulation of the two-tone flux-modulation protocol. Population dynamics from the effective model is compared with that from the full time-dependent model. The solid curves are obtained from the engineered effective Hamiltonian in the effective model, while the dashed curves are obtained from the explicitly modulated Hamiltonian in the full model. Other parameters used are $\Delta^{(1)}=\kappa/200$, $\Delta^{(2)}=-50\kappa$. The modulation indices are chosen $A_{1}/\omega_{d1}=0.2$ and $A_{2}/\omega_{d2}=0.9$. For a representative decay rate $\kappa/2\pi=50~\mathrm{MHz}$, the modulation frequencies are chosen with $\omega_{d1}/2\pi=500~\mathrm{MHz}$ and $\omega_{d2}/2\pi=900~\mathrm{MHz}$.}}
	\label{fig:A1}
\end{figure}
We also provide a numerical simulation of the explicitly modulated Hamiltonian to test the proposed modulation protocol. As shown in Fig.~\ref{fig:A1}, the dynamics of the full time-dependent Hamiltonian in Eq.~(\ref{H_rot}) agrees well with the engineered effective Hamiltonian in Eq.~(\ref{H_eng}). The remaining fast oscillations in the full model are induced by the high-frequency drives and can be averaged out in the slow timescale. The numerical results confirm that the engineered couplings between qubit and connecting modes together with the residual detunings can be generated from the proposed two-tone flux modulation.
\subsection{Generalization to two qubits}	
Finally, we extend the above derivation to the two qubits case. When each qubit $Q_{m}$ ($m=L,R$) is independently modulated by its own set of drive parameters, the slow dynamics is governed by the generalized Hamiltonian in Eq.~(\ref{H_eng}):
\begin{equation}
	\begin{aligned}
		\hat{H}_{\text{eng}}&=-\sum_{n=1}^2\Delta^{(n)}\hat{c}^{(n)\dagger} \hat{c}^{(n)} \\
		+& \sum_{n=1}^2 \left[\left( g_{L}^{(n)}\hat{\sigma}_{L}^+ + g_{R}^{(n)}\hat{\sigma}_{R}^+\right) \hat{c}^{(n)} + \text{H.c.} \right],
	\end{aligned}
	\label{H_eng2q}
\end{equation}
with the complex engineered coupling coefficients:
\begin{equation}
	\begin{aligned}
		g_{L/R}^{(1)}&=\lambda_{L/R}^{(1)} J_1\left(\frac{A_{L/R,1}}{\omega_{d1}}\right) J_0\left(\frac{A_{L/R,2}}{\omega_{d2}}\right) e^{i\tilde{\psi}_{L/R,1}}  \\
		g_{L/R}^{(2)}&=\lambda_{L/R}^{(2)} J_0\left(\frac{A_{L/R,1}}{\omega_{d1}}\right) J_1\left(\frac{A_{L/R,2}}{\omega_{d2}}\right) e^{i\tilde{\psi}_{L/R,2}}.
	\end{aligned}
\end{equation}
Eq.~(\ref{H_eng2q}) is the engineered effective Hamiltonian used in Eq.~(\ref{H_prime}) of the main text.
	
Consequently, the flux modulation scheme can directly produce the complex coupling coefficients $g_{L/R}^{(n)}$ including both amplitudes and phases with high flexibility.
\end{color}

\section{\label{sec:appendixB}The parameters used in our work}
To clarify the consistency between the decoherence rates used in Fig.~(\ref{fig:decoherence}) and the Hamiltonian parameters assumed throughout the paper, we explicitly calculate and list a experimentally feasible set of relevant parameters in two tables. While Table~\ref{tab:decoher-params} summarizes representative experimentally feasible values in physical units. And Table~\ref{tab:numerical-params} shows the dimensionless parameters used in the numerical simulations, normalized by the resonator decay rate $\kappa$.

\begin{table}[t]
	\caption{Representative experimentally feasible parameter scales used in the manuscript.}
	\label{tab:decoher-params}
	\begin{ruledtabular}
		\begin{tabular}{lll}
			Parameter & Value &  \\ \hline
			$\kappa/2\pi$ & $50~\mathrm{MHz}$ & engineered resonator loss \\
			$\gamma_{1,2}/2\pi$ & $10~\mathrm{kHz}$ & $\mathrm{T}_1 \approx 100~\mu\mathrm{s}$ \\
			$\gamma_\phi/2\pi$ & $30~\mathrm{kHz}$ & $\mathrm{T}_\phi \approx 33~\mu\mathrm{s}$
		\end{tabular}
	\end{ruledtabular}
\end{table}

\begin{table*}[t]
	\caption{Dimensionless parameters (in units of $\kappa$) used in the numerical simulations.}
	\label{tab:numerical-params}
	\begin{ruledtabular}
		\begin{tabular}{c c c c c}
			\multirow{2}{*}{Parameter} & \multicolumn{2}{c}{Set I} & \multicolumn{2}{c}{Set II} \\
			& n=1 & n=2 & n=1 & n=2 \\ \hline 
			$\Delta^{(n)}/\kappa$                & $\Delta^{(1)}/\kappa=\frac{1}{2}$ & $\Delta^{(2)}/\kappa=-\frac{1}{2}$  & $\Delta^{(1)}/\kappa=\frac{1}{200}$ & $\Delta^{(2)}/\kappa=-50$ \\
			$|\Delta^{(n)}+i\kappa/2|/\kappa$    & 0.707 & 0.707 & 0.500 & 50.002 \\
			$|g_{L/R}^{(n)}|/\kappa$             & 0.084 & 0.084 & 0.071 & 0.707
		\end{tabular}
	\end{ruledtabular}
\end{table*}

From Table~\ref{tab:decoher-params}, the qubit relaxation and dephasing rates are chosen as $\gamma_{1,2}/2\pi=10~\mathrm{kHz}$ and $\gamma_\phi/2\pi=30~\mathrm{kHz}$, corresponding to transmon lifetimes $T_1$ around $100~\mu\mathrm{s}$. In particular, for engineered decay rate $\kappa/2\pi=50~\mathrm{MHz}$, the ratio $\gamma/\kappa=2\times 10^{-4}$ used in Fig.~(\ref{fig:decoherence}) is even smaller than $10^{-3}\kappa$, confirming that the numerical simulations are performed in a conservative regime.

\begin{table*}[t]
	\caption{\textcolor[rgb]{0,0,0}{Representative parameters for the implementation of two-tone flux modulation.}}
	\label{tab:modulation-params}
	\begin{ruledtabular}
		\begin{tabular}{c c c c c}
			\multirow{2}{*}{Parameter} & \multicolumn{2}{c}{Set I} & \multicolumn{2}{c}{Set II} \\
			& $n=1$ & $n=2$ & $n=1$ & $n=2$ \\ \hline
			$\omega_{dn}/2\pi$ 
			& $500~\mathrm{MHz}$ 
			& $900~\mathrm{MHz}$  
			& $500~\mathrm{MHz}$ 
			& $900~\mathrm{MHz}$ \\
			
			$A_{L/R,n}/\omega_{dn}$ 
			& $1$ 
			& $1$ 
			& $0.2$ 
			& $0.9$ \\
			
			$\lambda_{L/R}^{(n)}/2\pi$ 
			& $\sim 12.5~\mathrm{MHz}$ 
			& $\sim 12.5~\mathrm{MHz}$ 
			& $\sim 44~\mathrm{MHz}$
			& $\sim 87~\mathrm{MHz}$ 
		\end{tabular}
	\end{ruledtabular}
\end{table*}

\begin{color}[rgb]{0,0,0}
We also summarize in Table~\ref{tab:modulation-params} one representative implementation of the two-tone flux modulation for each parameter set considered in the main text. The purpose of Table~\ref{tab:modulation-params} is not to define unique operating points, but to illustrate explicitly that the target modulation frequencies and engineered couplings used in the main text can be realized in experimentally reasonable parameters by choosing moderate dimensionless modulation indices $A_{L/R,n}/\omega_{dn}$, together with feasible bare couplings $\lambda_{L/R}^{(n)}$. 

For Set I, the target engineered parameters are $\Delta^{(1)}/2\pi=25~\mathrm{MHz}$, $\Delta^{(2)}/2\pi=-25~\mathrm{MHz}$, and $|g_{L/R}^{(n)}|/2\pi\sim4.2~\mathrm{MHz}$, as shown in Table~\ref{tab:numerical-params}. For instance, using modulation indices $A_{L/R,1}/\omega_{d1}=A_{L/R,2}/\omega_{d2}=1$ gives the relevant Bessel functions $J_0(1)\approx0.765$, $J_1(1)\approx0.440$, and $J_2(1)\approx0.115$, indicating that the first order sideband is appreciable and the higher order contributions can be ignored. In this case, the target bare couplings $\lambda_{L/R}^{(n)}/2\pi\sim12.5~\mathrm{MHz}$ and the drive frequencies can be chosen to be $\omega_{d1}/2\pi=500~\mathrm{MHz}$ and $\omega_{d2}/2\pi=900~\mathrm{MHz}$. For Set II, we use the modulation indices $A_{L/R,1}/\omega_{d1}=0.2$ and $A_{L/R,2}/\omega_{d2}=0.9$, where $J_0(0.2)\approx0.990$, $J_1(0.2)\approx0.100$, and $J_2(0.2)\approx0.005$ together with $J_0(0.9)\approx0.808$, $J_1(0.9)\approx0.406$, and $J_2(0.9)\approx0.094$. This implies that the higher order terms remain strongly suppressed. Therefore, the choice is sufficient to support the sideband coupling without making the modulation into an extreme Bessel function regime. The corresponding bare couplings ranging from around $12.5~\mathrm{MHz}$ to $87~\mathrm{MHz}$ are also consistent with coupling strength realized in current circuit QED platforms. Moreover, the low-frequency modulation tones are compatible with an arbitrary waveform generator (AWG), while the $\mathrm{GHz}$ modulation tone can be generated by a microwave source applied through the flux line~\cite{Zhou2021}. Therefore, Table~\ref{tab:modulation-params} provides an experimentally feasible parameters of the tunable engineered couplings considered in the main text.
\end{color}

\begin{acknowledgments}
P.B.L. is supported by the National Natural Science Foundation of China under Grants No. 12375018 and No. W2411002.
\end{acknowledgments}

%

\end{document}